\documentclass[a4paper, amsfonts, amssymb, amsmath, reprint, showkeys, nofootinbib,twocolumn,dvipsnames]{revtex4-2}
\usepackage[english]{babel}
\usepackage[utf8]{inputenc}
\usepackage[colorinlistoftodos, color=green!40, prependcaption]{todonotes}
\usepackage{amsthm}
\usepackage{mathtools}
\usepackage{physics}
\usepackage{xcolor}
\usepackage{graphicx}
\usepackage[left=13mm,right=13mm,top=25mm,columnsep=15pt]{geometry} 
\usepackage{adjustbox}
\usepackage{placeins}
\usepackage[T1]{fontenc}
\usepackage{lipsum}
\usepackage{csquotes}
\usepackage{ulem}
\usepackage{subfigure}
\usepackage{stackengine}
\usepackage{placeins}
\usepackage{array}
\usepackage[pdftex, pdftitle={Article}, pdfauthor={Author}]{hyperref} 
\usepackage{soul}
\usepackage{dsfont}
\usepackage{mathrsfs,amsmath}
\begin{document}

\title{Experimental sample-efficient and device-independent GHZ state certification}

\author{L. dos Santos Martins}
    \email[Correspondence email address:]{laura.dos-santos-martins@lip6.fr}
    \affiliation{Sorbonne Université, CNRS, LIP6, 4 Place Jussieu, Paris F-75005, France}

\author{N. Laurent-Puig}
    \email[Correspondence email address:]{nicolas.laurent-puig@lip6.fr}
    \affiliation{Sorbonne Université, CNRS, LIP6, 4 Place Jussieu, Paris F-75005, France}

\author{I. \v{S}upi\'c}
    \affiliation{Sorbonne Université, CNRS, LIP6, 4 Place Jussieu, Paris F-75005, France}

\author{D. Markham}
    \affiliation{Sorbonne Université, CNRS, LIP6, 4 Place Jussieu, Paris F-75005, France}
    
\author{E. Diamanti}
    \affiliation{Sorbonne Université, CNRS, LIP6, 4 Place Jussieu, Paris F-75005, France}

\date{\today} 

\begin{abstract}
The certification of quantum resources is a critical tool in the development of quantum information processing. In particular, quantum state verification is a fundamental building block for communication and computation applications, determining whether the involved parties can trust the resources at hand or whether the application should be aborted. Self-testing methods have been used to tackle such verification tasks in a device-independent (DI) setting. However, these approaches commonly consider the limit of large (asymptotic), identically and independently distributed (IID) samples, which weakens the DI claim and poses serious challenges to their experimental implementation. Here we overcome these challenges by adopting a theoretical protocol enabling the certification of quantum states in the few-copies and non-IID regime and by leveraging a high-fidelity multipartite entangled photon source. This allows us to show the efficient and device-independent certification of a single copy of a four-qubit GHZ state that can readily be used for the robust and reliable implementation of quantum information tasks.

\end{abstract}

\maketitle
The certification of entangled quantum states is one of the most important quantum information primitives, as entangled states are very often the difficult resource for quantum information applications, so that their certification can in turn certify the reliable running of the associated application~\cite{PhysRevLett.81.5932,eisert2020quantum}. Examples of such resources are cluster states as the resource for universal measurement based quantum computation~\cite{raussendorf2001one}, stabiliser states for error correction~\cite{gottesman1997stabilizer} and GHZ states, which are resources for anonymous communication~\cite{christandl2005quantum}, quantum metrology~\cite{giovannetti2011advances}, leader election~\cite{tani2012exact} and more. 

When entangled states are intended for use for a given application, their certification should meet three key requirements. 
First, it must output a state that can be used for a given application, without additional assumptions. Given that measurements in quantum theory are destructive and irreversible, it is crucial to certify the quality of quantum states without destroying them. 
Second, certification should ideally not rely on complete trust in the measurement devices, as faulty assumptions about their structure could compromise security, implying it should be, at least to some degree, device-independent. 
Lastly, the certification process must account for possible memory effects, thereby avoiding the assumption that the source produces independent and identical copies in every round of the experiment. 
In addition to the requirements imposed by the applications, certification should also prioritize efficiency in terms of both energy consumption and time investment. 
Indeed, quantum resources are inherently costly, underscoring the significance of verifying quantum systems using the fewest possible samples or measurements while still achieving robust confidence in the results. 
This not only minimizes the time, cost, and computational resources needed for verification but also mitigates the potential introduction of errors by external factors, narrowing the time window during which errors may occur. 

Bell nonlocality, aside from being a fundamentally nonclassical phenomenon used to invalidate locally-causal theories, offers an elegant solution for device-independent certification~\cite{brunner2014bell}. This approach enables certification despite lacking control over the measurement devices, with self-testing results pinpointing specific quantum experiments based solely on observed measurement correlations~\cite{vsupic2020self}. While traditionally expressed as mathematical theorems linking quantum setups with correlation probabilities obtained through infinite repetitions, recent advancements demonstrate the practical utility of such self-testing results in finite regimes, enabling protocols for sample-efficient device-independent certification of quantum states without the need for the independent and identical distribution (IID) assumption~\cite{PRXQuantum.3.010317}.

Such device-independent certification protocols pose significant experimental challenges and hence have been largely unexplored in practice. 
Recent experiments have certified states when measurement devices are trusted (i.e., not DI)~\cite{PhysRevLett.125.030506}, even when some parties act dishonestly~\cite{McCutcheon2016}. In the DI setting there have been experiments robustly self-testing states~\cite{PhysRevLett.127.230503,Xu:22}, or verifying entanglement properties~\cite{PhysRevLett.129.190503}, but these assume IID and are only valid in the large copy (asymptotic) limit. 
No experiment so far satisfies our three requirements for the certification of entangled quantum states.
Here we experimentally demonstrate, for the first time to our knowledge, the DI certification of a single copy of a four-partite GHZ state, completely free of the IID assumption. Our demonstration relies upon and expands the sample-efficient protocol of~\cite{PRXQuantum.3.010317} and leverages the characteristics of a high-performance multipartite entangled photon source~\cite{martins2024realizingcompacthighfidelitytelecomwavelength}. We analyze the protocol in terms of the achieved certified optimized fidelity measure and show that our implementation opens the way to carrying out efficiently and reliably quantum information tasks.

\begin{figure}
\centering
\includegraphics[width=0.7\columnwidth]{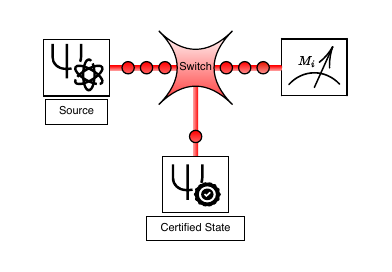}
\caption{\textbf{DI quantum state certification scheme.} The source generates $N$ states, out of which $N-1$ are used as a verification set and are measured according to Bell nonlocality tests. The outcomes resulting from these measurements are used to evaluate if the remaining unmeasured sample is certified to be close to the target state.
}
\label{fig:scheme}
\end{figure}

\vspace{5mm}
\noindent \textbf{A fully device-independent scenario}\\
\noindent The objective of our certification protocol is to quantitatively assess the proximity of a state, $\sigma_c$, generated by an uncharacterized source to the target state $\ket{GHZ}=(\ket{HHHH}+\ket{VVVV})/\sqrt{2}$, without direct measurement. Ideally, the source should consistently produce multiple copies of the state $\ket{GHZ}$. However, due to its uncharacterized nature, it is possible that, over $N$ rounds, the produced state, $\sigma^N$, may exhibit correlations or even entanglement across rounds. In our protocol, based on~\cite{PRXQuantum.3.010317}, we perform measurements over $N-1$ rounds and use the obtained results to estimate the proximity of an unmeasured copy to the target state, as illustrated in Fig.~\ref{fig:scheme}. Operating in a device-independent scenario, where measurements are uncharacterized and conform to a Bell scenario, we only have access to input-output correlations. For this reason, our approach involves testing the violation of a Bell inequality, which self-tests the target state: a high Bell violation implies that all $N-1$ measured copies are close to the target state. Given the random selection of the unmeasured copy, we can infer with high confidence that it is also close to the target state.

The fact that we operate in a DI scenario leads to a few caveats in our certification protocol. Unlike in many other approaches to quantum certification tasks (see e.g. \cite{PhysRevLett.123.260504}), the fidelity between states cannot serve as a standard metric in this setting. Under DI conditions, our protocol is limited to certifying states up to local isometries, at best. Therefore, to address the uncertainty inherent in treating all measurement devices as black boxes, we propose using extractability as an appropriate metric, as suggested in~\cite{PhysRevLett.117.070402}. Extractability represents fidelity optimized over all possible local isometries. A high extractability indicates the presence of an isometry capable of aligning the measured state closely with the target state.  As all other copies are measured, the extractability of the unmeasured copy is estimated conditionally on the outcomes of the performed measurements. In this sense the certified extractability is conditional, or, in other words, it characterizes the conditional state:
\begin{equation}
\begin{aligned}
    &\tilde{\sigma}_c=\frac{1}{p_{1,\ldots,c-1,c+1,\ldots,N}}\\&\operatorname{Tr}_{1, \ldots, c-1, c+1, \ldots, N}
    [(\otimes^{N\textbackslash\{c\}}_{k=1} M_{\mathbf{o}_k \mid \mathbf{i}_k}) \sigma^{N}],
\end{aligned}
\end{equation}

\noindent where $\sigma^N$ is the state over $N$ rounds, $M_{\mathbf{o}_k \mid \mathbf{i}_k}$ represents the measurement performed on the $k$th copy giving outcome $\mathbf{o}_k$, and $p_{1, \ldots, c-1,c+1,\ldots,N}=\operatorname{Tr}[(\otimes^{N\textbackslash\{c\}}_{k=1} M_{\mathbf{o}_k \mid \mathbf{i}_k}) \sigma^{N}]$. This approach guarantees that the conditional state of the unmeasured copy is independent from all the other copies produced in the measurement rounds, which allows us to define the appropriate figure of merit for a fully DI quantum state certification protocol and formally express its final goal. We wish to claim, with a confidence level $1-\delta$, whether the extractability of the conditional state, $\tilde{\sigma}_c$, from the target state, $\ket{GHZ}$, is bigger than some value $1-\eta$, with $\eta \in [0,1]$, which can be written as:
\begin{equation}
\Xi(\tilde{\sigma}_c,\ket{GHZ}) = \textup{max}_{\Phi}\mathscr{F}(\Phi[\tilde{\sigma}_c],\ket{GHZ}) \geq 1-\eta,
\label{eq:extractability}
\end{equation}

\noindent where $\Phi$ is an arbitrary local isometry and the fidelity of a state $\sigma$ with respect to the target state $\ket{\psi}$ is defined as $\mathscr{F}(\sigma, \ket{\psi})=\bra{\psi}\sigma\ket{\psi}$.
\begin{figure*}[!t]
\centering
\begin{subfigure}
\centering
\includegraphics[scale=0.359]{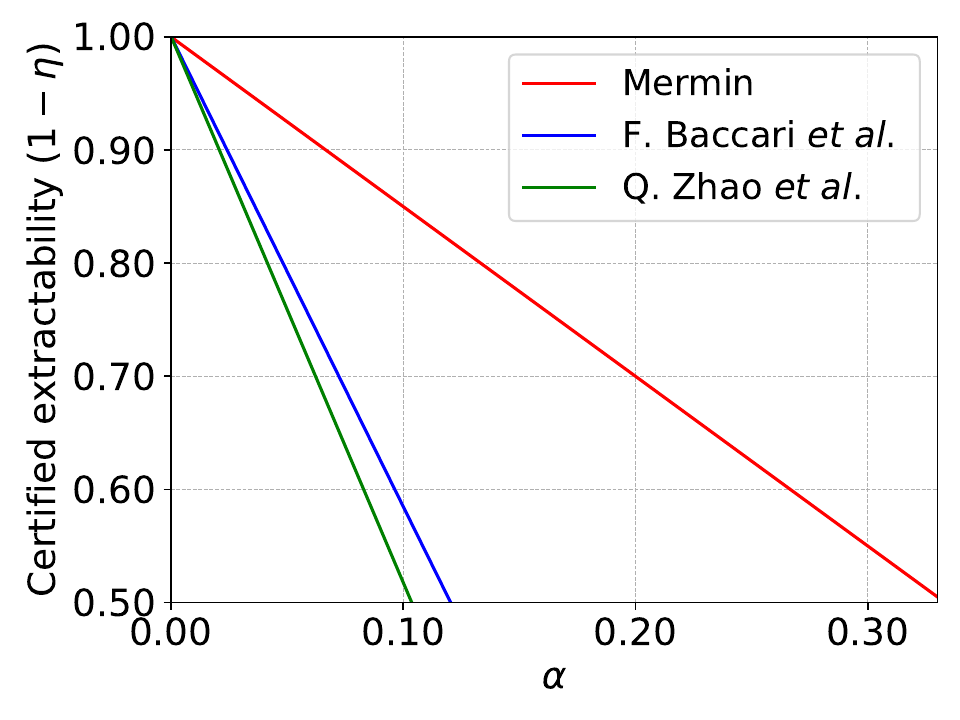}
\end{subfigure}
\begin{subfigure}
\centering
\includegraphics[scale=0.359]{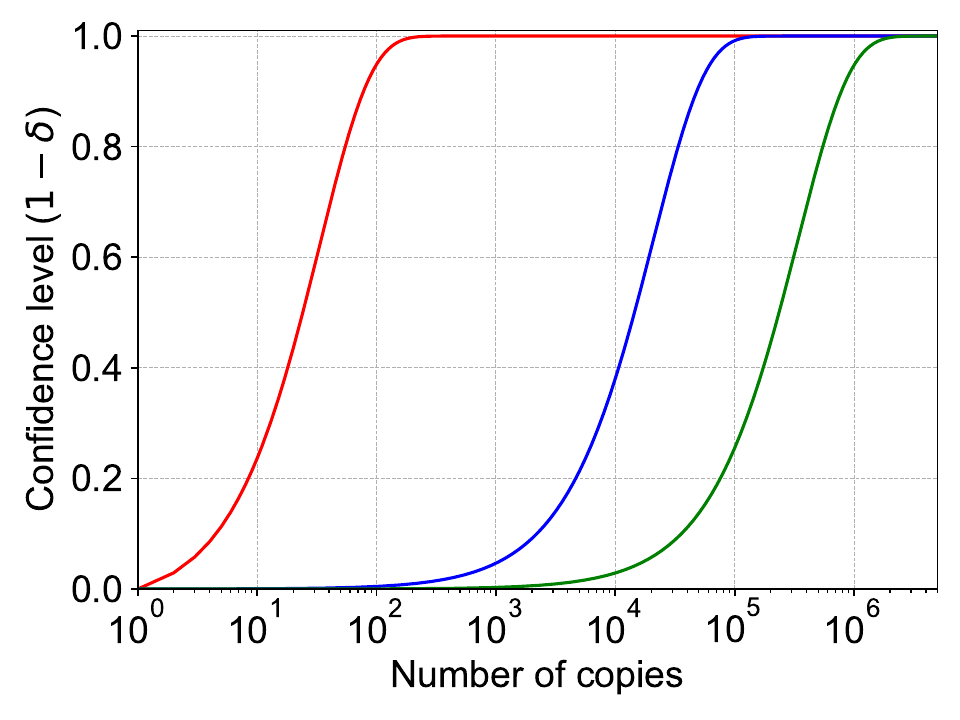}
\end{subfigure}
\begin{subfigure}
\centering
\includegraphics[scale=0.359]{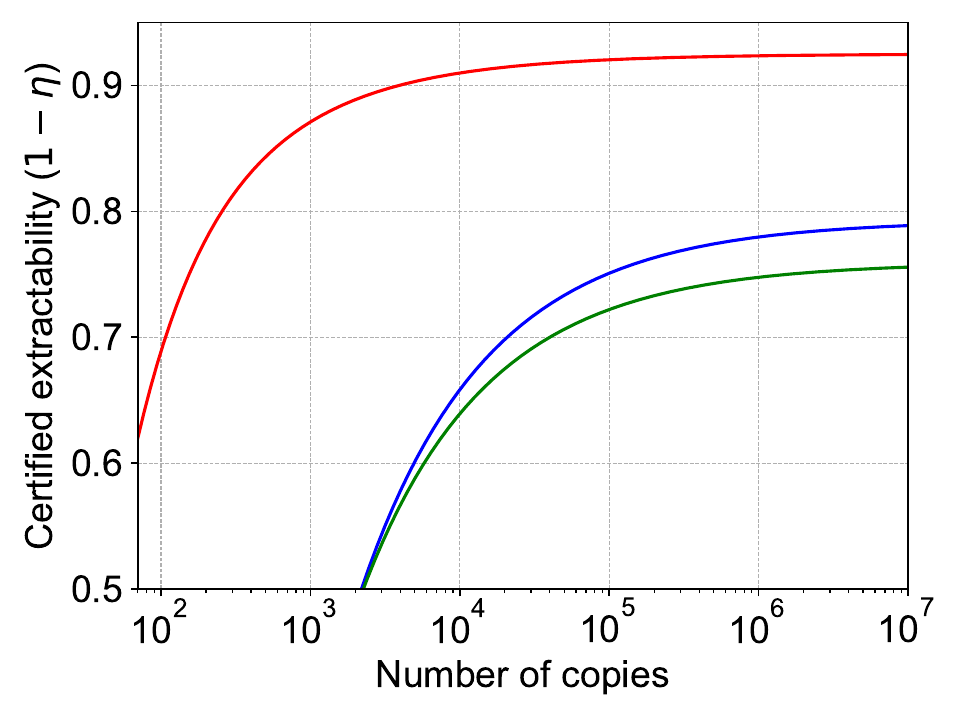}
\end{subfigure}
\caption{Performance of three distinct Bell operators in the certification of a four-qubit GHZ state: green and blue lines refer to operators proposed in~\cite{PhysRevLett.124.020402} and \cite{PhysRevLett.127.230503} (see also Eqs.~(\ref{eq:baccari_ineq}) and (\ref{eq:Bell Inequality}) in the Methods), respectively, for robust self-testing purposes; red line corresponds to the Mermin operator (Eq.~(\ref{eq:Mermin})). \textbf{Left}: Maximum certified extractability, $1-\eta$, as a function of the certified quantum state noise, $\alpha$, for a confidence level of $1-\delta=0.99$, and setting $\epsilon_1=\epsilon_2$, which is only possible in the limit of $N \rightarrow \infty$; \textbf{Middle}: Confidence level, $1-\delta$, as a function of the total number of copies required for the certification task, $N$, for a certified infidelity of $\eta = 0.25$ and for $\alpha=0.05$. \textbf{Right}: Certified extractability, $1-\eta$, as a function of the total number of copies required for the certification task, $N$, for a fixed $1-\delta = 0.99$, $\alpha=0.05$.}
\label{fig:simulations}
\end{figure*}

In order to properly estimate the extractability of the GHZ state from the unmeasured copy, we must first carefully choose a Bell inequality that self-tests the target state. In other words, the selected Bell inequality should be maximally violated only by the $\ket{GHZ}$ state (up to local isometries). This selection determines the Bell test to which the copies will be subjected during the measurement rounds. Furthermore, we can rely on robust self-testing statements based on a Bell inequality to establish a lower bound on the extractability of the underlying quantum state, $\Xi(\tilde{\sigma}_c,\ket{GHZ})$, from the observed Bell violation, $\beta$. Moving from a general self-testing framework to a well-defined certification protocol, it is useful to reframe the scenario as a nonlocal game derived from the Bell inequality. In this context, after establishing the appropriate winning and losing outcomes (where winning corresponds to those outcomes that contribute to violation of the Bell inequality), only the target state (up to local isometries) achieves the optimal quantum winning probability, $p_{QM}$. Although self-testing statements are typically designed for IID sources, we can leverage the robustness statement to determine the maximal winning probability for states with limited extractability. 

After the copies are measured and given a score for their performance in all $N-1$ rounds (they score one if they get a winning result, zero if not), we can add them up to determine the overall score, $N_{win}$, and deduce the resulting verification pass rate, $P = N_{win} / (N-1)$. 

For a given desired extractability $1-\eta$ (Eq.~(\ref{eq:extractability})), and confidence level $1-\delta$, the number of copies required is determined by two parameters $\epsilon_1,\epsilon_2$ which are related to the violation of the Bell inequality in the ideal case. In the protocol, $\epsilon_1$ fixes the required pass rate: our claim on extractability holds when $P\geq p_{QM}-\epsilon_1$. This is then related to the desired extractability through $\epsilon_2$, via $c\eta = \epsilon_2 > \epsilon_1$, where $c$ is a constant coming from self-testing, linking the extractability to Bell violation (see~\cite{PRXQuantum.3.010317} and Supplementary Information). The role of $\epsilon_2$ is to allow for a gap in the requested pass rate and the desired extractability so that our goal can be achieved for finite $N$. 
The requested number of copies $N$ is chosen so that the following is satisfied
\begin{equation}
\delta \leq \left(\frac{1}{N} + \frac{N-1}{N}e^{D(p_1\mid \mid p_2)}\right)^N,
\label{eq:N_certification}
\end{equation}

\noindent where $D(a||b) = a\log(a/b) + (1-a)\log[(1-a)/(1-b)]$ is the Kullback-Leibler divergence and $p_i=p_{QM}-\epsilon_i$; see Supplementary Information for a detailed description of the protocol.

\begin{figure}[!htbp]
 \includegraphics[width=\columnwidth]{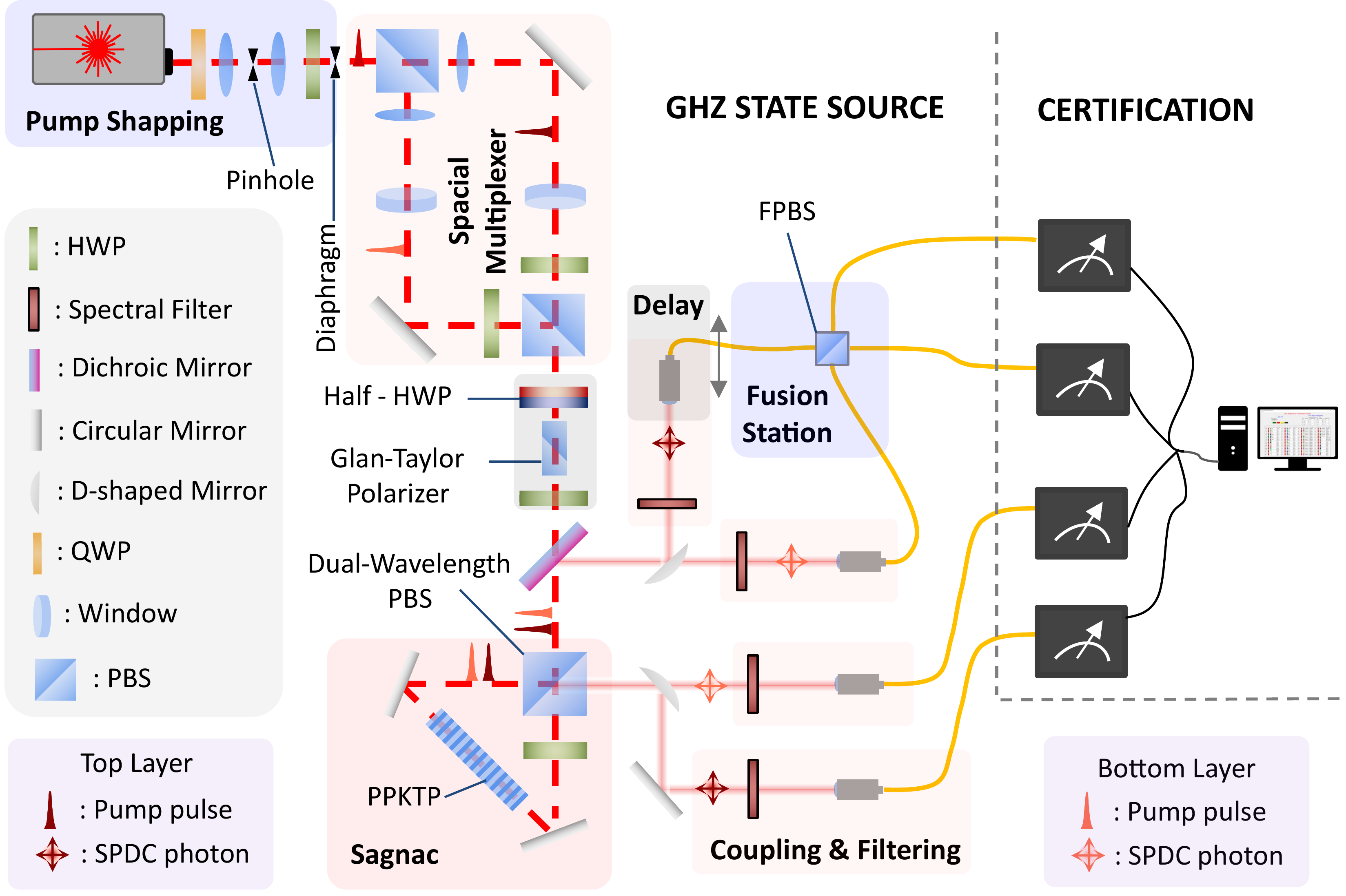}
        \caption{Experimental setup. \textbf{Laser pump:} A Ti:Sapphire laser with an average power of 3.4~W emits 2~ps pulses at a wavelength of 775~nm with a repetition rate of 76~MHz. \textbf{Spatial mode shaping:} The spatial mode of the laser is shaped into a Gaussian profile. \textbf{Spatial multiplexer:} The pump pulses are split into two parallel beams: the top and bottom layers, horizontally and vertically polarized, respectively. \textbf{Polarization shaping:} Both layers are diagonally polarized, in order to maximize the fidelity of both layers output states to the Bell state. \textbf{Sagnac interferometer:} Photon pairs are probabilistically generated via type-II SPDC in a ppKTP crystal (30mm-long, 46.2é $\mu$m poling period) and entangled in polarization in the Sagnac loop, resulting in the output state $(\ket{H}_s\ket{V}_i+e^{i\theta}\ket{V}_s\ket{H}_i)/\sqrt{2}$. \textbf{Coupling and Filtering:} After filtering the single photons with a dichroic mirror, 1100~nm long pass filters and 1.3~nm ultra-narrowband filters, the bottom layer photons are reflected on half-circle shaped mirrors, while the top layer photons are transmitted over them. The photons are coupled to single-mode fibers with 12~mm focal lens. \textbf{Fusion Station:} The mechanical delay on the bottom layer idler photon is finely tuned such that both idler photons (from the top and bottom layers) arrive simultaneously to the fibered polarization beam splitter (FPBS). If each one of them is transmitted to different outputs of the FPBS, and conditioned on fourfold coincidences, a GHZ state $(\ket{HHHH}+e^{i\delta}\ket{VVVV})/\sqrt{2}$ is generated. \textbf{Certification:} In each round, each of the four black-boxes receives an input, which determines the measurement setting and outputs the resulting outcome. Given the winning and losing outcomes we can calculate the overall score, $N_{win}$, over $N-1$ rounds and use the results to certify the proximity of a randomly selected copy to the target state, $\ket{GHZ}$.}
        \label{fig:Setup}
\end{figure}

\begin{figure*}[!htbp]
\centering
{\includegraphics[width=0.495\textwidth]{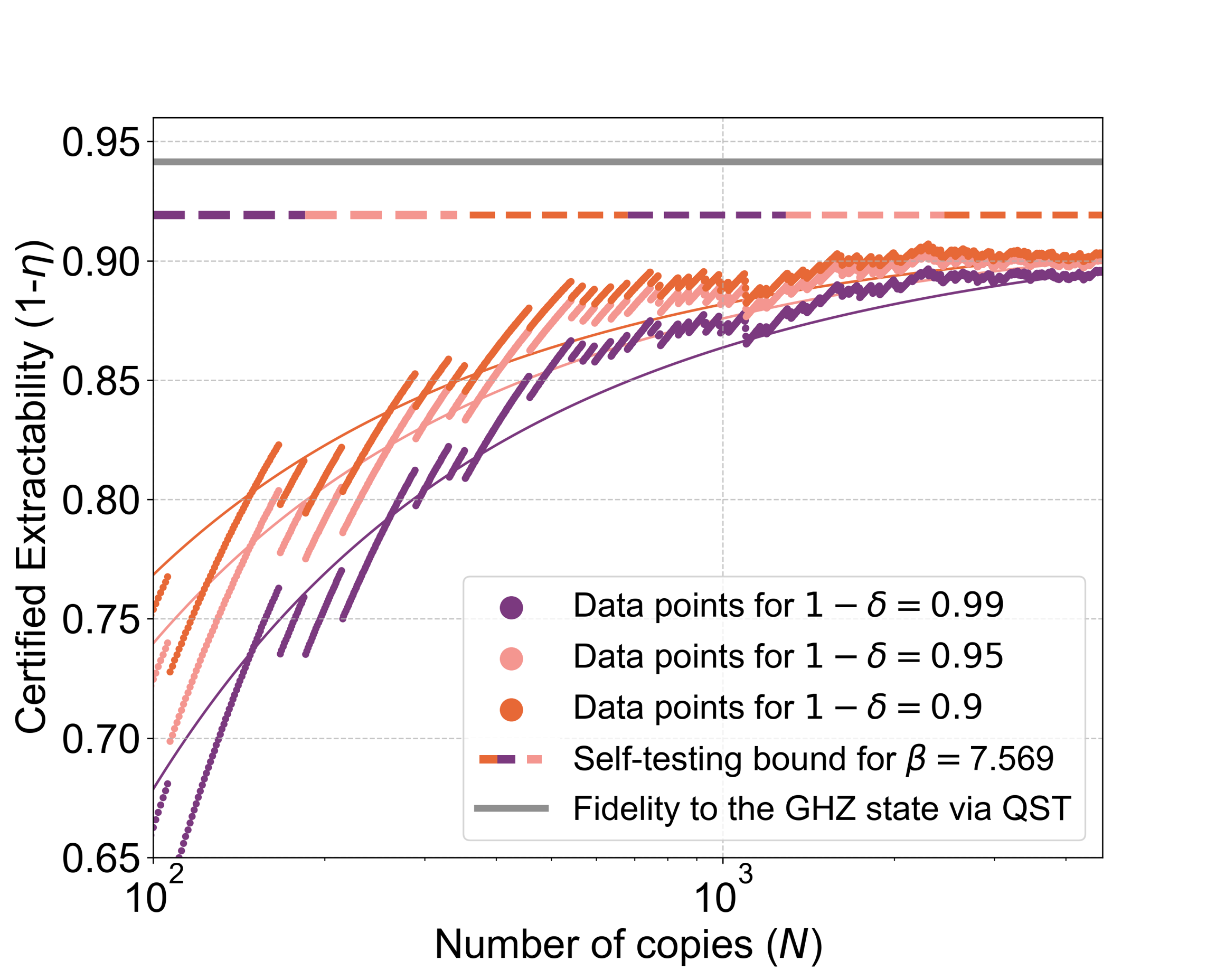}}
\begin{subfigure}
\centering
{\includegraphics[width=0.495\textwidth]{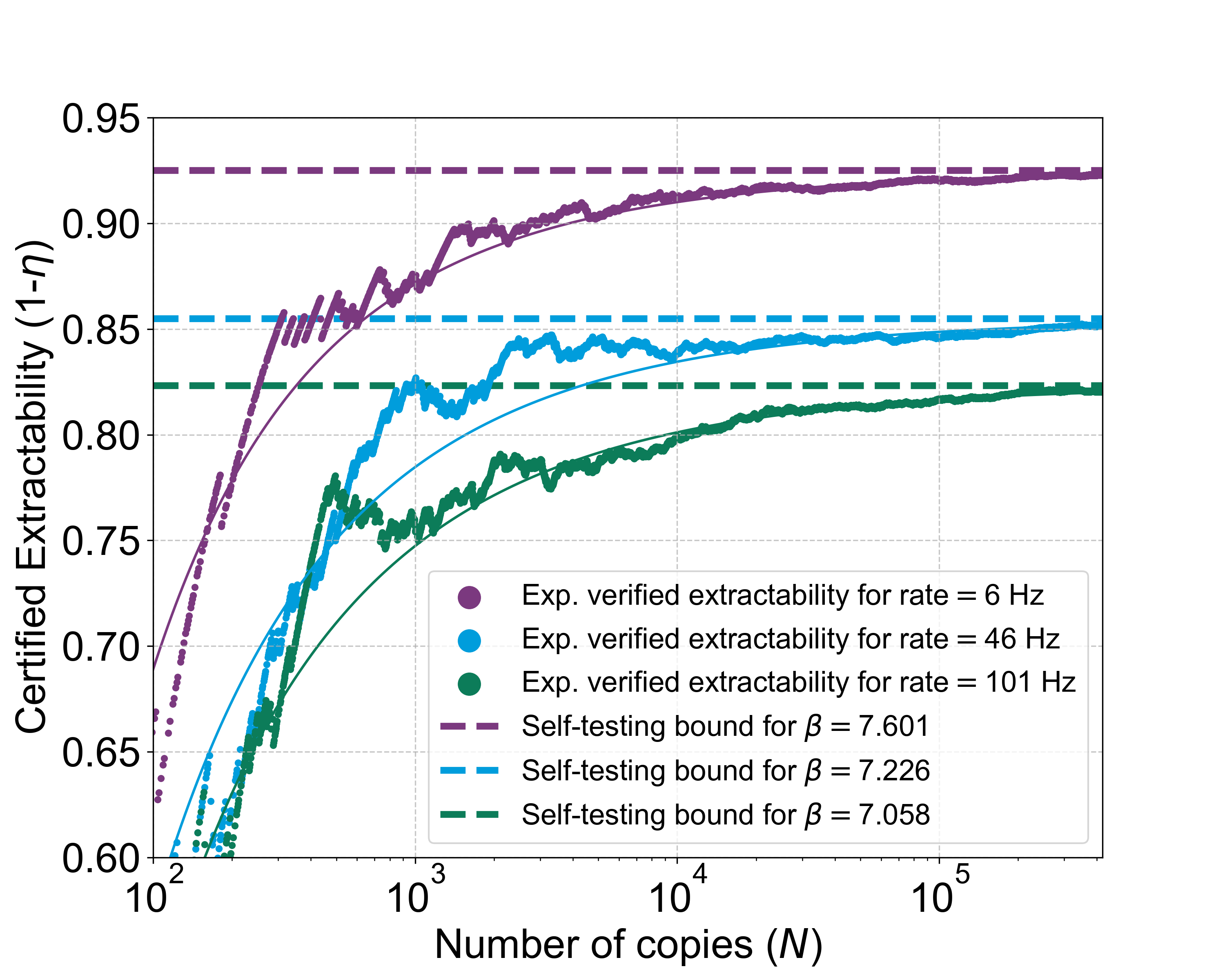}}
\end{subfigure}
\caption{Experimental certified extractability of a four-qubit GHZ state with respect to the total number of samples, $N$. Full simulation curves consider the winning probability threshold to be $p_1=N_{win}/(N^{max}-1)$. The dashed horizontal lines refer to the self-testing bounds, determined by the measured Mermin violations, $\beta$, representing the maximum obtainable certified extractability in the limit of $N \rightarrow \infty$. \textbf{Left:} Fully DI protocol implementation for different confidence levels, $1-\delta$. Each classical input is associated with one single output. Data set taken for a pump power of 240~mW, yielding an average state generation rate of 6~Hz. The full horizontal grey line refers to the fidelity estimated via Quantum State Tomography $(\mathscr{F}=94.15\%)$ and shows the correctness of the protocol; \textbf{Right:} Protocol implementation for different rates: $6$~Hz (purple), $46$~Hz (blue) and $101$~Hz (green). Each input is associated with multiple outcomes recorded within an acquisition time of 15~seconds. The confidence level is fixed to $1-\delta=0.99$.}
\label{fig:Verifed_Fidelity}
\end{figure*}

\vspace{5mm}
\noindent \textbf{Choosing the Bell measurement operator}\\
\noindent As explained above, a crucial step for experimentally demonstrating the certification of a GHZ state is to select the most appropriate Bell operator. To this end, we studied the performance of three different candidates: two Bell operators, proven to have very tight self-testing bounds in terms of robustness~\cite{PhysRevLett.124.020402, PhysRevLett.127.230503} (see Methods, Eqs.~(\ref{eq:baccari_ineq}) and (\ref{eq:Bell Inequality})) and the following Mermin-like operator~\cite{PhysRevA.77.062106}:
\begin{equation}
\begin{split} \label{eq:Mermin}
\mathbf{B}_{Mermin} & = A_{0}B_{0}C_{0}D_{0} - A_{1}B_{1}C_{0}D_{0}\\
&-A_{1}B_{0}C_{1}D_{0} -A_{1}B_{0}C_{0}D_{1} \\
&+A_{1}B_{1}C_{1}D_{1} -A_{0}B_{1}C_{1}D_{0} \\
&-A_{0}B_{1}C_{0}D_{1} -A_{0}B_{0}C_{1}D_{1},
\end{split}
\end{equation}

\noindent where $A_0,B_0,C_0,D_0 = X$ and $A_1,B_1,C_1,D_1 = Y$. The motivation behind this choice is the fact that its quantum bound, $\beta_{Q}$, saturates the algebraic bound, $\beta_{algebraic}$, leading to a maximum success probability of $p_{QM}=1$. However, to our knowledge, the only self-testing bound for a Mermin inequality existing in the literature is restricted to a tripartite system~\cite{PhysRevLett.117.070402}. For this reason, we computed for this work a self-testing bound for the four-partite case, relying on the numerical method described in~\cite{PhysRevLett.124.020402} (see
Supplementary Information for more details).

To compare the different Bell operators fairly, we consider a quantum state described as a statistical mixture of a GHZ state and white noise, mathematically expressed as $\rho=(1-\alpha)\rho_{GHZ}+\frac{\alpha}{16}\mathds{1}$. We calculate the pass rate (as a probability in this case) $P$, and subsequently fix the winning probability threshold, $p_1=P$, for each operator, as a function of the noise characterizing the state we want to certify, $\alpha$. This allows us to study the behaviour of the remaining parameters, captured in inequality~(\ref{eq:N_certification}). For this analysis, three main figures of merit stand out for their significance: the maximum extractability one can certify, how many samples one needs to measure in order to complete the protocol and the confidence level associated with the results. In Fig.~\ref{fig:simulations} we analyse the behaviour of the different operators from the perspective of the parameters mentioned above (further details are given in the Methods). In all three plots, it is clear that the Mermin operator outperforms the others, providing not only a significant advantage in terms of sample efficiency, by almost two orders of magnitude, but an overall better performance, regarding how close we can certify a state with respect to the target state. It is therefore the one we use to device-independently certify a quantum state.

\vspace{5mm}
\noindent \textbf{Experimental results}\\
To experimentally demonstrate the fully DI certification of a quantum state, we use a compact and high-performance four-party GHZ state source, based on spontaneous parametric down conversion (SPDC) in a layered-Sagnac interferometer configuration~\cite{martins2024realizingcompacthighfidelitytelecomwavelength} (see Fig.~\ref{fig:Setup}). Once the states are generated, we transmit each of the four photons to the measurement apparatus and run the protocol described above. After the protocol is successfully completed, we use all the recorded measurement outcomes, except for one corresponding to a single copy randomly selected, to calculate the pass rate $P$. After setting the desired confidence level, $1 - \delta$, we use the total number of copies, $N$, to numerically invert inequality~(\ref{eq:N_certification}) and compute the solution for the maximum certified extractability that fulfills the condition $\epsilon_2 > \epsilon_1$ (a detailed description of the data acquisition and its analysis is given in the Methods).

The results are shown in Fig.~\ref{fig:Verifed_Fidelity}. The certified extractability with respect to the total number of copies follows the overall expected behaviour outlined by the simulation (full) curves, apart from standard experimental fluctuations. The plot on the left illustrates that, by tuning the confidence level, we can reduce the number of samples required to achieve the same certification level. However, this trade-off vanishes for large samples, as the certified extractability always converges to the self-testing bound. These results show, for the first time to our knowledge, the experimental device-independent certification of a four-qubit GHZ state with an extractability of $\Xi(\tilde{\sigma}_c, \ket{GHZ}) \geq 0.896$, for a total of $4643$ verified samples and a confidence level of $1-\delta=0.99$, in a non-IID scenario. This demonstration is only possible due to the high-fidelity states produced with our experimental setup, yielding an average success probability of $P=0.973$. 

It is clear that the collection of additional samples would allow us to get closer to the self-testing bound of $\Xi_{max}(\tilde{\sigma}_c,\ket{GHZ}) \geq 0.919$, assuming the average passing probability of $P$ would remain unchanged. However, a significant increase of the acquired statistics can be experimentally challenging due to the accumulation of the recovery time associated with updating each black-box's setting. This motivates the consideration of a less conservative trust scenario, in which, instead of a one-to-one input-to-output correspondence, we record multiple outcomes within the 15 seconds acquisition time associated with each classical input. Using post-processing techniques, we can decompose the results obtained within the acquisition window into multiple near-single-shot measurements, mimicking the recording of a single output for each input. A randomization of the decomposed events provides the opportunity to simulate the implementation of the protocol for a larger sample, using experimental data (see Methods for more details about the data acquisition and analysis). The resulting data set, shown in Fig.~\ref{fig:Verifed_Fidelity} (right), is further explored by testing different state generation rates - ranging from 6~Hz to 101~Hz - which result in varying degrees of high-order SPDC emissions, consequently affecting the certified extractability - ranging from 0.923 to 0.0820, respectively. This analysis highlights the range of capabilities in which our multipartite entanglement source can operate and shows a clear convergence of the certified extractability towards the self-testing bound, which is only possible due to the drastic increase of copies to $N\sim 4\times10^5$.  Although this approach does not strictly follow the protocol, it shows that, as long as the stability of the setup is maintained, it is possible to saturate the self-testing bound.

\vspace{5mm}
\noindent \textbf{Discussion}\\
It is worth noting that, although the Mermin operator provides the best certification results among those we analysed, this does not mean they cannot be further improved. In fact, it is possible that the self-testing bound we found is not optimal in terms of robustness, i.e., that a tighter bound exists. Furthermore, there might exist operators, other than the ones we analysed, yielding a more favorable combination of robust self-testing bound parameters and maximum probability of winning, $p_{QM}$. Since the GHZ state certification protocol heavily relies on such parameters, this suggests that the same experimental data can potentially produce even better results. Additionally, while the experimental setup in Fig.~\ref{fig:Setup} is suitable for the purpose of demonstrating the certification of a quantum state, integrating this protocol as a subroutine in a quantum information task would likely require optical switches to guide the certified copies to their intended purpose. Moreover, our primary limitation in recording a large sample, while guaranteeing a one-to-one input-to-output correspondence, is the active control of the Mermin settings. More specifically, the experimental realisation of each black-box involves controlling the rotation of waveplates with mechanical motors. Adjusting their configuration requires waiting for their response time before measuring, which is accumulated over all classical inputs, leads to time-consuming implementations. Alternatively, a passive choice mechanism could accelerate the protocol execution, making it more efficient.

This work reinforces the validity of the fully DI certification of quantum states as a valuable fundamental resource for a wide range of quantum information applications. We emphasize the practicality of our protocol in providing a rigorous framework in which a finite number of samples can yield meaningful results, without further assuming identical and independent distribution for all produced copies. Furthermore, it is instructive to observe the impact the number of samples has on the robustness of our results. This is particularly clear when comparing the purple data points in the two plots in Fig.~\ref{fig:Verifed_Fidelity} - for a similar passing probability and the same confidence level, the certified extractability increases by $3\%$ with ten times more samples. In other words, while the theory is able to characterize the few-copies regime, demanding confidence levels and extractability requirements can only be achieved for relatively large samples. 

In general, the ability to experimentally demonstrate the fully DI certification of such a high extractability level paves the way to the reliable and robust use of quantum information systems in practical, real-world settings.

\vspace{5mm}
\noindent\textbf{Note.} At the time of finalising this work, we became aware of parallel and independent work on experimental quantum state certification, also submitted to the arXiv today: M. Antesberger \textit{et al.}, ``Efficient and Device-Independent Active Quantum State Certification'' (2024).

\vspace{5mm}
\noindent\textbf{Acknowledgments.} We thank Uta Isabella Meyer and Henrique Silvério for fruitful discussions and technical support. We acknowledge financial support from the European Union’s Horizon 2020 framework programme under the Marie Sklodowska Curie innovation training network project AppQInfo, Grant No. 956071 (LdSM), the Horizon Europe research and innovation programme under the project QSNP, Grant No. 101114043 (ED), the European
Research Council Starting Grant QUSCO, Grant No. 758911 (NLP, ED), the PEPR integrated projects QCommTestbed, ANR-22-PETQ-0011 (ED) and EPiQ, ANR-22-PETQ-0007 (IS, DM), and the HQI project, ANR-22-PNCQ-000 (DM), which are part of Plan France 2030.
\vspace{5mm}

\noindent\textbf{{METHODS}}\\
\noindent\textbf{{Self-testing bounds}}\label{sec:Self-TestingBound}\\
The self-testing bound plays a pivotal role in our implementation of the quantum state certification protocol, as it impacts not only the sample efficiency but also the lower bound for the certified extractability. For this purpose, as mentioned in the main text, we consider two options, introduced in~\cite{PhysRevLett.127.230503,PhysRevLett.124.020402}, alongside the Mermin operator (Eq.~(\ref{eq:Mermin})). The Bell operator, derived from F. Baccari~\textit{et al.}~\cite{PhysRevLett.124.020402} holds significant potential for its tight self-testing bound and it can be written as:
\begin{equation}
\begin{split}
        &\mathbf{B}_{F.Baccari} =  3(A_{0}B_{0}C_{0}D_{0} + A_{1}B_{0}C_{0}D_{0}) + \\
        &(A_{0}B_{1} -A_{1}B_{1}) + (A_{0}C_{1} -A_{1}C_{1}) +(A_{0}D_{1} -A_{1}D_{1}).
\end{split}
\label{eq:baccari_ineq}
\end{equation}

\noindent The subsequent operator, originating from Q. Zhao~\textit{et al.}~\cite{PhysRevLett.127.230503}, offers comparable advantages, it was demonstrated to be even tighter than the first one, and it is defined as:
\begin{equation}
    \begin{split}\label{eq:Bell Inequality}
    &\mathbf{B}_{Q.Zhao_{1}} = (A_{0}+A_{1})B_{1}C_{1}D_{1} \\
    &+(A_{0}-A_{1})B_{0} + B_{0}C_{0} + B_{0}D_{0}.\\\\
    \end{split}
\end{equation}
\noindent For GHZ states, the optimal quantum bounds - for both operators mentioned above - can be achieved by taking $A_0=\frac{X + Z}{\sqrt{2}}$, $A_1=\frac{X - Z}{\sqrt{2}}$ and $B_{i},C_{i},D_{i}$ defined as $X$ ($Z$) for $i=0$ ($i=1$). Lastly, we take the Mermin operator, defined in Eq.~(\ref{eq:Mermin}), due to its optimal maximum success probability of $p_{QM}=1$.

\begin{figure}[!htbp]
\includegraphics[width=\columnwidth]{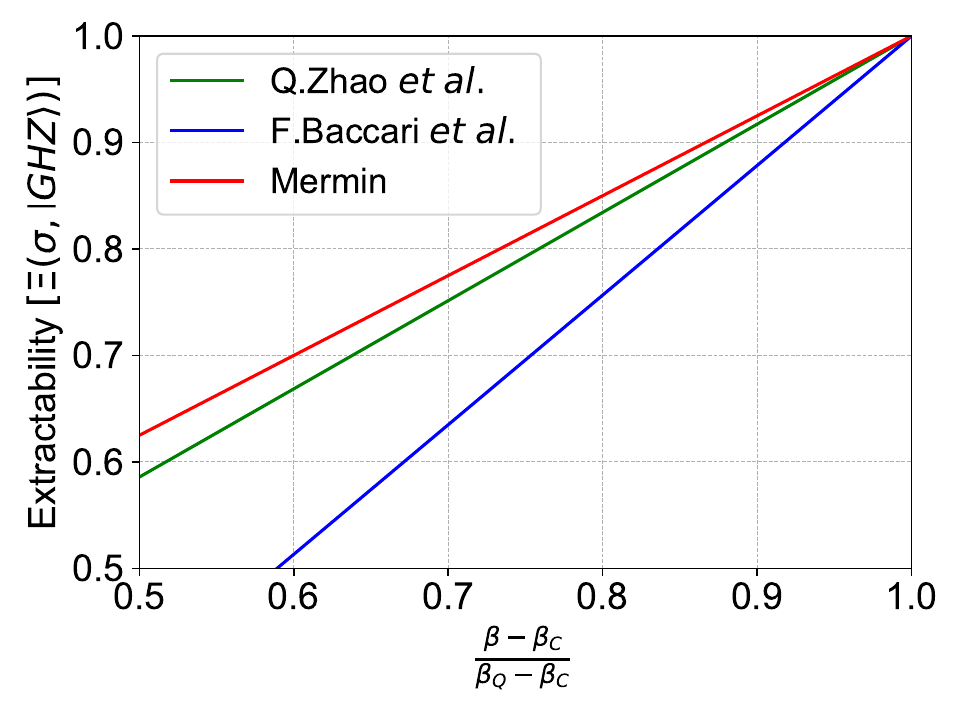}
\caption{Lower bound of the extractability from the target GHZ state as a function of the normalized violation $\frac{\beta - \beta_C}{\beta_Q - \beta_C}$. Green and blue lines refer to inequalities proposed in~\cite{PhysRevLett.124.020402} and \cite{PhysRevLett.127.230503}, respectively Eqs.~(\ref{eq:baccari_ineq}) and (\ref{eq:Bell Inequality}), and the red line corresponds to the Mermin inequality (Eq.~(\ref{eq:Mermin})).}
\label{fig:self-testing}
\end{figure}

In order to compare the different options, we consider the robust self-testing statement to be of the form (see Supplementary Information for more details):
\begin{equation}
    \Xi(\sigma, \ket{GHZ}) \geq s\beta+\mu,
\label{eq:self-test_statement}
\end{equation}
\noindent where $s, \mu \in \mathbb{R}$, for all states $\sigma$ achieving a violation greater than $\beta$. The lower bound of each inequality, determined by the values of $s$ and $\mu$ (see Table~\ref{tab:self-testing}), is illustrated in Fig.~\ref{fig:self-testing}.

\begin{table}[h]
\centering
\begin{tabular}{|c|c|c|c|c|}
\hline
\textbf{4-qubit GHZ} & \boldmath$s$ & \boldmath$\mu$ & \boldmath$\beta_Q$ & \boldmath$\beta_C$ \\
\hline
$\mathbf{B}_{F.Baccari}$ & 0.4897 & -3.1552 & $6\sqrt{2}$ &  6\\
\hline
$\mathbf{B}_{Q.Zhao_{1}}$ & 1 & $-1 - 2\sqrt{2}$  & $2\sqrt2+2$ & 4\\
\hline
$\mathbf{B}_{Mermin}$ & 0.1875 & -0.5 & 8 & 4 \\
\hline
\end{tabular}
\caption{Numerical self-testing bound parameters for the three different operators considered for the device-independent (DI) quantum state certification protocol.}
\label{tab:self-testing}
\end{table}

The Mermin operator displays the tightest bound from a robust self-testing perspective, suggesting that it is likely the most appropriate operator for the certification protocol. However, since the lower bound on the extractability provided by the DI quantum state certification protocol also depends on the maximum probability of winning the nonlocal game with a quantum strategy, $p_{QM}$, we observe that the advantage of the Mermin operator becomes even more predominant than one would think by simply comparing the self-testing bounds (see Fig.~\ref{fig:simulations}). This indicates that the robust self-testing analysis does not contain all the necessary information for the choice of the most favorable operator for the DI certification of a quantum state.

\vspace{5mm}
\noindent\textbf{{Data collection and analysis}}\label{sec:AnalysisMethod}\\
\noindent We conducted the experiment for three different pump power settings, associated with different state generation rates: 6 Hz, 46 Hz and 101 Hz. For each of these configurations, we collected more than $4\times10^5$ states over a fixed acquisition window of 15 seconds per randomly selected classical input.

To implement the fully DI certification protocol, we must guarantee the precise isolation of one single random event from the whole sample. One possibility could be to employ an optical switch. While this method allows for the selection of some states, accurately discerning the presence of one and only one state would prove challenging due to the post-selected nature of states produced by SPDC and their inherent high-order emissions. Alternatively, post-processing analysis techniques are a viable solution. Using a high-performance time tagger, we can record the precise time-stamps associated with each detected event. With this capability, we can \textit{replay} the full experiment and decompose each of the 15-second acquisitions into multiple ultra-short measurements, such that each of them records, on average, one single output. Two different methods are taken to analyse the resulting data. First, we take each 15-second window and randomly select one out of the full set of recorded outcomes, guaranteeing a true one-to-one input-to-output correspondence in the nonlocal game (Fig.~\ref{fig:Verifed_Fidelity} - left). While this approach discards a significant portion of the recorded events and limits the number of measured samples, $N$, to the total number of randomly generated inputs throughout the experiment, it reflects a faithful implementation of the protocol. Alternatively, we consider the full dataset resulting from the decomposition of each classical input into as many inputs as the total number of recorded outputs within the 15-second acquisition window (Fig.~\ref{fig:Verifed_Fidelity} - right). As mentioned in the main text, while this last approach does not strictly follow the protocol, it allows to use the full data set to simulate a considerably larger sample. For both approaches, we can use a random number generator to select a single copy to be certified and, consequently, excluded from the verification analysis. It is worth noting that high-order emissions hamper the isolation of individual single events. Each time one of these is selected to be certified, we restart the random selection until we discard one and only one copy for each classical input.
\bibliography{references}

\begin{thebibliography}{10}

\bibitem{PhysRevLett.81.5932}
H.-J. Briegel, W.~D\"ur, J.~I. Cirac, and P.~Zoller, ``Quantum repeaters: The role of imperfect local operations in quantum communication,'' {\em Phys. Rev. Lett.}, vol.~81, pp.~5932--5935, Dec 1998.

\bibitem{eisert2020quantum}
J.~Eisert, D.~Hangleiter, N.~Walk, I.~Roth, D.~Markham, R.~Parekh, U.~Chabaud, and E.~Kashefi, ``Quantum certification and benchmarking,'' {\em Nature Reviews Physics}, vol.~2, no.~7, pp.~382--390, 2020.

\bibitem{raussendorf2001one}
R.~Raussendorf and H.~J. Briegel, ``A one-way quantum computer,'' {\em Physical review letters}, vol.~86, no.~22, p.~5188, 2001.

\bibitem{gottesman1997stabilizer}
D.~Gottesman, {\em Stabilizer codes and quantum error correction}.
\newblock California Institute of Technology, 1997.

\bibitem{christandl2005quantum}
M.~Christandl and S.~Wehner, ``Quantum anonymous transmissions,'' in {\em International conference on the theory and application of cryptology and information security}, pp.~217--235, Springer, 2005.

\bibitem{giovannetti2011advances}
V.~Giovannetti, S.~Lloyd, and L.~Maccone, ``Advances in quantum metrology,'' {\em Nature photonics}, vol.~5, no.~4, pp.~222--229, 2011.

\bibitem{tani2012exact}
S.~Tani, H.~Kobayashi, and K.~Matsumoto, ``Exact quantum algorithms for the leader election problem,'' {\em ACM Transactions on Computation Theory (TOCT)}, vol.~4, no.~1, pp.~1--24, 2012.

\bibitem{brunner2014bell}
N.~Brunner, D.~Cavalcanti, S.~Pironio, V.~Scarani, and S.~Wehner, ``Bell nonlocality,'' {\em Reviews of modern physics}, vol.~86, no.~2, pp.~419--478, 2014.

\bibitem{vsupic2020self}
I.~{\v{S}}upi{\'c} and J.~Bowles, ``Self-testing of quantum systems: a review,'' {\em Quantum}, vol.~4, p.~337, 2020.

\bibitem{PRXQuantum.3.010317}
A.~Go\ifmmode~\check{c}\else \v{c}\fi{}anin, I.~\ifmmode \check{S}\else \v{S}\fi{}upi\ifmmode~\acute{c}\else \'{c}\fi{}, and B.~Daki\ifmmode~\acute{c}\else \'{c}\fi{}, ``Sample-efficient device-independent quantum state verification and certification,'' {\em PRX Quantum}, vol.~3, p.~010317, Feb 2022.

\bibitem{PhysRevLett.125.030506}
W.-H. Zhang, C.~Zhang, Z.~Chen, X.-X. Peng, X.-Y. Xu, P.~Yin, S.~Yu, X.-J. Ye, Y.-J. Han, J.-S. Xu, G.~Chen, C.-F. Li, and G.-C. Guo, ``Experimental optimal verification of entangled states using local measurements,'' {\em Phys. Rev. Lett.}, vol.~125, p.~030506, Jul 2020.

\bibitem{McCutcheon2016}
W.~McCutcheon, A.~Pappa, B.~A. Bell, A.~McMillan, A.~Chailloux, T.~Lawson, M.~Mafu, D.~Markham, E.~Diamanti, I.~Kerenidis, J.~G. Rarity, and M.~S. Tame, ``Experimental verification of multipartite entanglement in quantum networks,'' {\em Nature Communications}, vol.~7, Nov. 2016.

\bibitem{PhysRevLett.127.230503}
D.~Wu, Q.~Zhao, X.-M. Gu, H.-S. Zhong, Y.~Zhou, L.-C. Peng, J.~Qin, Y.-H. Luo, K.~Chen, L.~Li, N.-L. Liu, C.-Y. Lu, and J.-W. Pan, ``Robust self-testing of multiparticle entanglement,'' {\em Phys. Rev. Lett.}, vol.~127, p.~230503, Dec 2021.

\bibitem{Xu:22}
J.-M. Xu, Q.~Zhou, Y.-X. Yang, Z.-M. Cheng, X.-Y. Xu, Z.-C. Ren, X.-L. Wang, and H.-T. Wang, ``Experimental self-testing for photonic graph states,'' {\em Opt. Express}, vol.~30, pp.~101--111, Jan 2022.

\bibitem{PhysRevLett.129.190503}
C.~Zhang, W.-H. Zhang, P.~Sekatski, J.-D. Bancal, M.~Zwerger, P.~Yin, G.-C. Li, X.-X. Peng, L.~Chen, Y.-J. Han, J.-S. Xu, Y.-F. Huang, G.~Chen, C.-F. Li, and G.-C. Guo, ``Certification of genuine multipartite entanglement with general and robust device-independent witnesses,'' {\em Phys. Rev. Lett.}, vol.~129, p.~190503, Oct 2022.

\bibitem{martins2024realizingcompacthighfidelitytelecomwavelength}
L.~dos Santos~Martins, N.~Laurent-Puig, P.~Lefebvre, S.~Neves, and E.~Diamanti, ``Realizing a compact, high-fidelity, telecom-wavelength source of multipartite entangled photons,'' 2024.

\bibitem{PhysRevLett.123.260504}
H.~Zhu and M.~Hayashi, ``Efficient verification of pure quantum states in the adversarial scenario,'' {\em Phys. Rev. Lett.}, vol.~123, p.~260504, Dec 2019.

\bibitem{PhysRevLett.117.070402}
J.~m.~k. Kaniewski, ``Analytic and nearly optimal self-testing bounds for the clauser-horne-shimony-holt and mermin inequalities,'' {\em Phys. Rev. Lett.}, vol.~117, p.~070402, Aug 2016.

\bibitem{PhysRevLett.124.020402}
F.~Baccari, R.~Augusiak, I.~\ifmmode \check{S}\else \v{S}\fi{}upi\ifmmode~\acute{c}\else \'{c}\fi{}, J.~Tura, and A.~Ac\'{\i}n, ``Scalable bell inequalities for qubit graph states and robust self-testing,'' {\em Phys. Rev. Lett.}, vol.~124, p.~020402, Jan 2020.

\bibitem{PhysRevA.77.062106}
A.~Cabello, O.~G\"uhne, and D.~Rodr\'{\i}guez, ``Mermin inequalities for perfect correlations,'' {\em Phys. Rev. A}, vol.~77, p.~062106, Jun 2008.

\bibitem{PhysRevA.73.012316}
T.~Kim, M.~Fiorentino, and F.~N.~C. Wong, ``Phase-stable source of polarization-entangled photons using a polarization sagnac interferometer,'' {\em Phys. Rev. A}, vol.~73, p.~012316, Jan 2006.

\bibitem{LR-QST}
B.~Qi, Z.~Hou, L.~Li, D.~Dong, and G.~X. .~G. Guo, ``Quantum state tomography via linear regression estimation,'' {\em Scientific Reports}, vol.~3, no.~3496, 2013.

\bibitem{PhysRevLett.108.070502}
J.~A. Smolin, J.~M. Gambetta, and G.~Smith, ``Efficient method for computing the maximum-likelihood quantum state from measurements with additive gaussian noise,'' {\em Phys. Rev. Lett.}, vol.~108, p.~070502, Feb 2012.

\end{thebibliography}
\bibliographystyle{ieeetr}
\appendix*
\clearpage
\section*{Supplementary Information} \label{sec:appendix}

\subsection{Detailed  protocol}\label{appendix:protocol}
In this section, we provide a complete and more general description for the estimation of the sample efficiency bound (Eq.~(\ref{eq:N_certification})) and the quantum state certification protocol adopted in this work and inspired by~\cite{PRXQuantum.3.010317}.

We start by considering a source that produces $N$ independent copies of a quantum system, denoted as $S = \{\sigma_1, \ldots, \sigma_N\}$, where $\sigma_i$ represents the $i$-th copy of a quantum state. For now, we assume the copies are independently distributed, but not identical, and then comment on the generalization. The purpose of quantum state certification is to quantitatively assess how close a set of states, $S_c$, is to the target state, $\ket{\psi}$. This can be done using the notion of average fidelity, $\bar F(S_c,\psi)$, where  \(S_c = \{\sigma_1, \ldots, \sigma_{N_c}\}\) denotes a set of \(N_c\) states. Our goal is to claim, with a confidence level $1-\delta$, whether the average fidelity of the set of samples, $S_c$, to the target state, $\ket{\psi}$, is bigger than some value $1-\eta$, with $\eta \in [0,1]$. This can be formally written as
\begin{equation} \bar F(S_c,\psi) = \frac{1}{N_c}\sum_{j=1}^{N_c}\bra\psi\sigma_j\ket\psi \geq 1-\eta.\end{equation}

\noindent However, in device-independent scenarios, local measurements are not characterized, or trusted, since all devices are treated as black boxes. In other words, some local isometries are undetectable, which renders state fidelity based verification methods impossible. To address this issue, we can leverage the concept of extractability, defined in the literature as the fidelity optimized across all possible isometries~\cite{PhysRevLett.117.070402}. More specifically, the extractability of the target state $\ket{\psi}$ from a state $\sigma_j$ is written as:
\begin{equation}
\Xi(\sigma_j,\psi)=\textup{max}_{\Phi}\mathscr{F}\{\textup{Tr}_{j}[\Phi(\sigma_j)],\ket{\psi}\},
\end{equation}
\noindent where $\Phi$ is an arbitrary local isometry. If we generalize this equation for the set $S_c$, we can re-define our goal with the average extractability: 
\begin{equation}
\bar\Xi(S_c,\psi) = \frac{1}{N_c}\sum_{j=1}^{N_c}\Xi(\sigma_j,\ket{\psi}) \geq 1-\eta.
\label{eq:general_goal}
\end{equation}

\noindent A high extractability implies the existence of an isometry that can bring the measured state close to the target state. In other words, we ensure that if we apply the inverse local isometry to any arbitrary measurement, the statistics obtained from the measured state will be close to those from the target state, which suggests that the extractability is the DI equivalent of the fidelity.

As mentioned above, so far, our analysis has presumed that the copies are uncorrelated with each other. However, this assumption can be misleading and compromise the protocol. Two different perspectives can be adopted to face this issue. We can restrict the protocol to the quantum state certification of one single copy and abandon the IID assumption, adopting the idea of conditional extractability, as detailed in the main text (see Eq.~(\ref{eq:extractability})). Alternatively, if we insist on certifying a set of states with more than one element ($N_c>1$), to our knowledge, as far as the theory is developed, we need to keep the assumption that all copies are uncorrelated with each other. In the latter case, it is useful to define $\mu=N_c/N$ as the fraction of certified samples. Independently of which option we follow, once the general goal of state certification is defined, we can move our focus towards the estimation of the average extractability. For this purpose, the proposed protocol adopts the form of a nonlocal game based on a Bell test. With this approach, once a Bell inequality that self-tests the target state, $\ket{\psi}$, is selected, we can use robust self-testing results to establish a lower bound on the average extractability as a function of the violation of the selected Bell inequality~\cite{PhysRevLett.117.070402}. More precisely, the robustness statement -  characterized by a constant $\tilde{c}$, itself dependent on the Bell operator under analysis - asserts that, for a given state, $\sigma$, achieving the Bell violation $\beta = \beta_Q - \eta/\tilde{c}$, where $\beta_Q$ is the quantum bound - defined as the maximal violation achievable by the correlations compatible with quantum theory - there exists an isometry, $\Phi$, such that Eq.~(\ref{eq:general_goal}) holds. The full protocol is detailed as follows:

\begin{enumerate}
    \item A source generates $N$ copies of a quantum state. Each copy, $\sigma_j$, is distributed over each spatially separated and non-communicating player, $k$.
    
    \item After all states were distributed, the verifier rolls an $N$-faced die, until $N_c=\mu N$ different outcomes are obtained, therefore determining the set of states, $S_c$, to be preserved and certified. The remaining states constitute the verification set, $S_v$, which will be measured.
    
    \item To decide the measurement setting for each copy, $j$, in the verification set, each player, $k$, will randomly generate an input, $i_{k,j}$, determining the measurement setting of their black-box, which will output a certain result $o_{k,j}$.
    
    \item After gathering the score of each copy from the verification sample, i.e., a winning output was obtained, given the received input, we can calculate the overall pass rate $P=N_{win}/N_v$.
\end{enumerate}

\noindent The success of the protocol can only be evaluated after we fix the desired extractability, characterized by $\eta$, which in turn, defines the lower bound of the average success probability of the whole sample $p_2=p_{QM} - \epsilon_2$, with $\epsilon_2 = c\eta$, $p_{QM}$ being the maximum probability of winning the nonlocal game with a given quantum strategy and $c=(2\tilde{c}\beta_{algebraic})^{-1}$ serving as a fundamental factor in establishing a link between the self-testing bound and the nonlocal game. Additionally, if we acknowledge that the success probability threshold of the verified sample, $p_{1} = p_{QM} -\epsilon_{1}$, needs to be larger than $p_2$, i.e., $\epsilon_1 < \epsilon_2$, we can define the condition that determines the successful certification of the remaining copy, expressed as $P \geq p_{1} = p_{QM} -\epsilon_{1}$. While studying the theoretical aspects of our work, we discovered an inconsistency between the violation of the Bell inequality and the winning probability of the nonlocal game, which eventually leads to the certification of an extractability higher than the self-testing bound. The origin of this disparity is in the definition of the constant $c$, lacking a factor of two in order to make a perfect translation of the Bell violation into a nonlocal game. For this reason, throughout the analysis of the experimental data, we considered the corrected expression $c=\frac{1}{2\tilde{c}\beta_{algebraic}}$. Depending on the specified parameters mentioned above, we can calculate the confidence level, associated with the certification process,
\begin{equation}
N \geq \frac{\ln \delta}{\ln(1-\mu + \mu e^{D(p_{\mathrm{QM}}-\epsilon_1\mid \mid p_{\mathrm{QM}}-\epsilon_2)})},
\label{eq:N_certification_general}
\end{equation}
\noindent where $D(a||b) = a\log(a/b) + (1-a)\log[(1-a)/(1-b)]$ is the Kullback-Leibler divergence.

\subsection{Robust self-testing bound}\label{appendix:self-testing-bound}
In this section, we detail the method used to estimate the robustness results of self-testing, associated with the 4-qubit Mermin operator (Eq.~(\ref{eq:Mermin})), shown in Table~\ref{tab:self-testing}. Following a similar approach to the one described in F. Baccari~\textit{et al.}~\cite{PhysRevLett.124.020402}, we recall that our goal is to find a lower bound for the extractability of the target state, $\ket{GHZ}$, from the measured state, $\sigma$, based on the violation, $\beta$, obtained with the Mermin operator, $\mathbf{B_{Mermin}}$. More concretely, we want to find $s, \mu \in \mathbb{R}$, such that:
\begin{equation}
     \Xi(\sigma, \ket{GHZ})\geq s\beta+\mu.
\end{equation}

\noindent If we recall that the extractability can be written as,
\begin{equation}
    \Xi(\sigma, \ket{GHZ}) =\underset{\Lambda=\Lambda_1\otimes\ldots\otimes\Lambda_4}{\max}\mathscr{F}\left(\Lambda(\sigma),\ket{GHZ}\right),
\label{eq:extractability_quantum_channel}
\end{equation}
\noindent where $\Lambda_i$ is the local channel on the $i$-th party and the fidelity is expressed as:
\begin{equation}
\begin{split}
&\mathscr{F}\left(\Lambda(\sigma),\ket{GHZ}\right)=\\ &\Tr\left[\sigma(\Lambda_1^\dag\otimes\ldots\otimes\Lambda_4^\dag)(\ket{GHZ}\bra{GHZ})\right],
\end{split}
\end{equation}
\noindent we can, equivalently, focus on finding the appropriate $s$ and $\mu$ parameters, such that, for some quantum channels, $\Lambda_i$, the following operator inequality holds:
\begin{equation}
\begin{split}
K &\vcentcolon= (\Lambda_1^\dag\otimes\ldots\otimes\Lambda_4^\dag)(\ket{GHZ}\bra{GHZ})\\
&\geq s \mathbf{B_{Mermin}} + \mu\mathds{1}.
\end{split}
\end{equation}
\noindent Since the measurement of $\mathbf{B_{Mermin}}$ only involves two dichotomic measurements per player, Jordan's lemma can be used to reduce the state to the $N$-qubit space, as mentioned in Ref.~\cite{PhysRevLett.124.020402}, allowing the parameterization of the local observables as,
\begin{equation}
\begin{split}
    A_i&=B_i=C_i=D_i\\
    &=\cos(\alpha_i)\sigma_++(-1)^i\sin(\alpha_i)\sigma_-,
\end{split}
\end{equation}
\noindent where $\sigma_+=(X+Y)/\sqrt{2}$ and $\sigma_-=(X-Y)/\sqrt{2}$ and $\alpha_i\in[0,\pi/2]$. Consequently, the operator $\mathbf{B_{Mermin}}(\vec{\alpha})$ can now be defined with respect to the angles $\alpha_i$. With the same intent in mind for $K$, we consider the following depolarizing channel, as suggested in Ref.~\cite{PhysRevLett.117.070402}:
\begin{equation}
    \Lambda_i(\alpha_i)=\frac{1+g(\alpha_i)}{2}\sigma+\frac{1-g(\alpha_i)}{2}\Gamma_i(\alpha_i)\sigma\Gamma_i(\alpha_i),
\end{equation}
\noindent where $g(\alpha_i) = (1 + \sqrt{2})(\sin \alpha_i + \cos \alpha_i - 1)$ and
\begin{equation}
\Gamma_i(\alpha_i)=\begin{cases}
    \sigma_+ & \text{for $\alpha_i \in [0,\pi/4]$} \\
    \sigma_- &\text{for $\alpha_i \in (\pi/4,\pi/2]$}
\end{cases}.
\end{equation}

\noindent After the operators' parameterization, we move our focus towards proving, for all possible $\alpha_i$, that the following inequality is satisfied:
\begin{equation}
    K(\alpha_0, \ldots, \alpha_4) \geq s \mathbf{B_{Mermin}} (\alpha_0, \ldots, \alpha_4)+\mu\mathds{1}
\end{equation}
\noindent for some $s,\mu \in \mathbb{R}$. To find the optimal parameters, we start by fixing $s$ and finding $\mu$ such that the extractability bound leads to 1 at the point of maximal violation, i.e., $\mu=1-s\beta_Q$.  For that combination of $s$ and $\mu$, we then check that the minimum eigenvalue of $K(\vec{\alpha})-s\mathbf{B_{Mermin}}(\vec{\alpha})-\mu\mathds{1}$ is larger or equal than 0, for all $\alpha_i$. If this condition is verified, we repeat the previous steps for a lower $s$. The optimal bound is determined by the minimum value of $s$ and corresponding $\mu$, satisfying the imposed condition.

\subsection{Experimental details}
\begin{figure}[!htbp]
\centering
\includegraphics[scale=0.25]{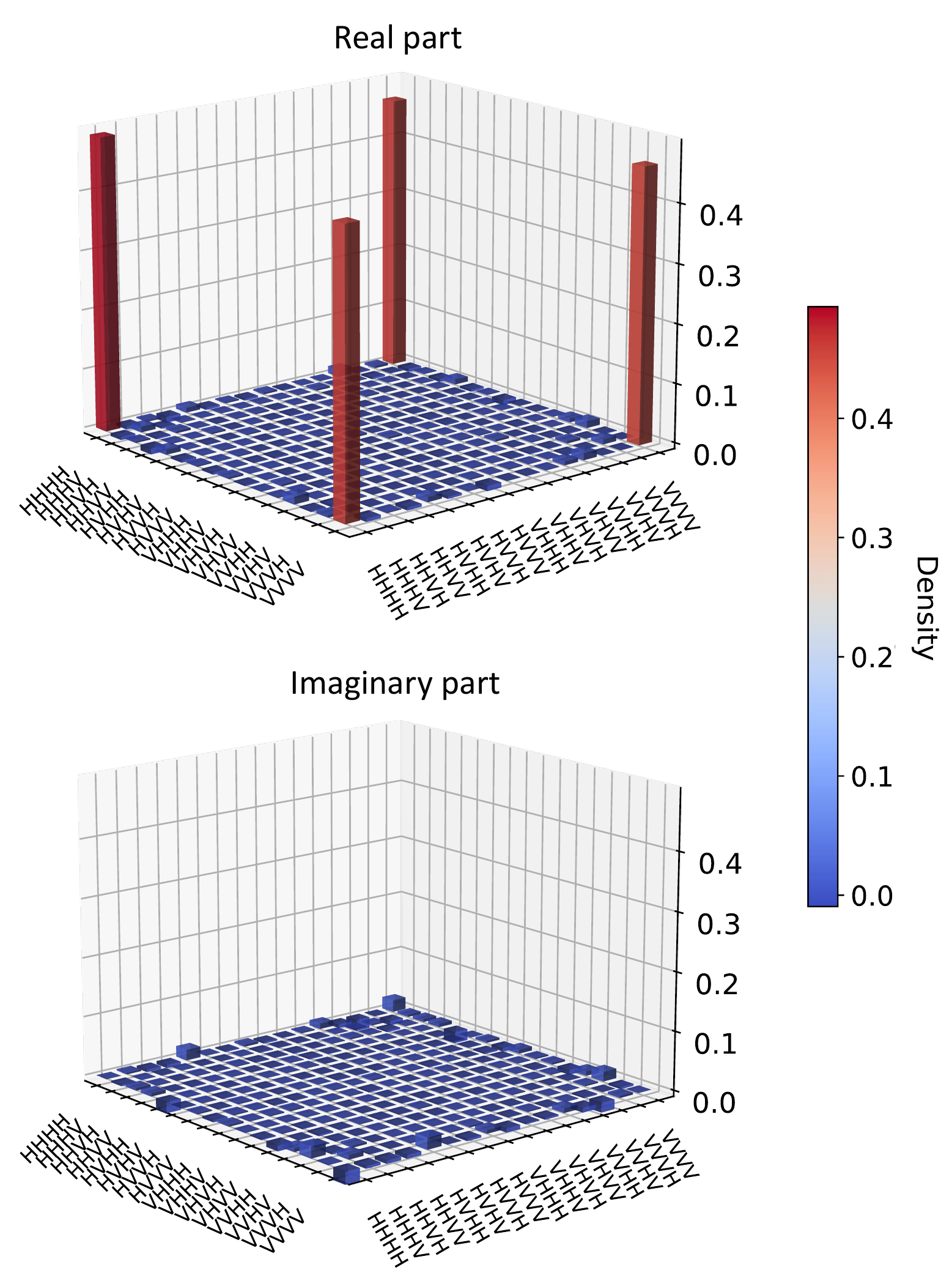}
 \caption[dm]{Experimental density matrix of the state produced with our source, estimated  via Quantum State Tomography, at 6.7~GHZ states per second. The fidelity with respect to the GHZ state is $(94.15 \pm 0.21)\%$ for an acquisition time of 150~s per basis.}
  \label{fig:dm}
\end{figure}

\noindent The polarization-entangled state used in our certification protocol is generated through the entanglement fusion of two Bell pairs. Each state is produced via type-II spontaneous parametric down conversion (SPDC) occurring within a periodically-poled KTP (ppKTP) crystal. To create the two Bell states, we split the pump into two parallel beams, using a spatial multiplexer, to pump the same ppKTP crystal in two different locations (top and bottom). Polarization entanglement is achieved by pumping the crystal from two opposing directions and then interfering the two paths using a Polarizing Beam Splitter (PBS), realized with a Sagnac interferometer~\cite{PhysRevA.73.012316}. As a result, we obtain the Bell state $\ket{\Phi} = (\ket{HV} + e^{i\theta}\ket{VH})/\sqrt{2}$, where $\theta$ is determined by the path difference between the two directions of propagation. We extract one photon from each pair and guide them to interfere on a Fiber Polarizing Beam Splitter (FPBS). Using a motorized delay stage, we finely adjust the temporal overlap of the interfering photons. We post-select the events resulting in each photon occupying a different spatial port of the FPBS. In other words, conditioned on fourfold coincidences, we entangle the two Bell pairs, thereby generating a GHZ state of the form $\ket{GHZ} = (\ket{HHHH} + e^{i\delta}\ket{VVVV})/\sqrt{2}$, where $\delta$ is determined by the $\theta$ of each Bell pair. With the aforementioned setup, we can generate a GHZ state up to local unitaries resulting from the propagation of the state in single-mode fibers. To specifically produce the state $\ket{GHZ}$, with $\delta=0$, we begin by performing Quantum State Tomography (QST) to precisely identify the state being generated. Subsequently, an optimization method is employed to determine the necessary local unitaries required to transform the state to the desired form. Using three sets of Quarter-Wave Plates (QWPs), Half-Wave Plates (HWPs), and Quarter-Wave Plates (QWPs), one for each of three out of the four photons, we apply those unitaries to achieve the target state. More details can be found in~\cite{martins2024realizingcompacthighfidelitytelecomwavelength}. Please note that the unitary compensation part of the setup is not shown explicitly in the quantum certification experimental setup illustrated in Fig.~\ref{fig:Setup}.\\

For a pump power of 240~mW, yielding a fourfold coincidence rate of $6.7$~Hz, we obtain a fidelity of $\mathscr{F}=\vert\bra{GHZ}\rho_{exp}\ket{GHZ}\vert^2 =(94.15\pm0.21)\%$ (see Fig.~\ref{fig:dm}). To evaluate the fidelity, we use Quantum State Tomography, based on linear regression~\cite{LR-QST} and fast maximum likelihood estimation~\cite{PhysRevLett.108.070502}. To assess the uncertainty associated with the reconstructed state, we employ the Monte Carlo method by sampling 500 times from Poissonian photon counting statistics and Gaussian QHP-HWP rotation angle distributions (which incorporates the systematic measurement basis error).
\end{document}